\documentclass[preprint,showpacs,preprintnumbers,amsmath,amssymb,nofootinbib]
{revtex4}
\usepackage{epsfig}
\begin{document}

\begin{flushright}
\end{flushright}


\newcommand{\be}{\begin{equation}}
\newcommand{\ee}{\end{equation}}
\newcommand{\bea}{\begin{eqnarray}}
\newcommand{\eea}{\end{eqnarray}}
\newcommand{\nn}{\nonumber}

\def\lb{\Lambda_b}
\def\ll{\Lambda}
\def\mb{m_{\Lambda_b}}
\def\ml{m_\Lambda}
\def\s1{\hat s}
\def\ds{\displaystyle}


\title{\large Study of FCNC mediated $Z$ boson effect in the
 semileptonic rare baryonic decays $\lb \to \ll~ l^+ l^- $ }
\author{A. K. Giri$^1$,
R. Mohanta$^2$ }
\affiliation{$1$ Department of Physics, National Tsing Hua University,
Hsinchu, Taiwan 300\\
$^2$ School of Physics, University of Hyderabad, Hyderabad - 500 046,
India}

\begin{abstract}
We study the effect of the FCNC mediated $Z$ boson in the rare
semileptonic baryonic decays $\lb \to \ll~ l^+ l^-$. We consider the 
model where the standard model fermion 
sector is extended by an extra vector like down quark, as a consequence
of which it allows CP violating  $Z$ mediated flavor changing
neutral current at the tree level. We find that due to this 
non-universal $Zbs $ coupling, branching ratios of the rare
semileptonic $\lb$ decays are enhanced reasonably from their 
corresponding standard model values and the zero point of the
forward backward asymmetry for $\lb \to \ll~ \mu^+ \mu^-$ is
shifted to the left. 
\end{abstract}

\maketitle

\section{Introduction}
In recent years B-physics (studies relating to particles containing
a bottom quark or antiquark) has been a very active area of research, both
experimentally as well as theoretically. Giant machines, named
B-factories, have been constructed and are being in operation now
to study the dynamics of these heavy particles. This in turn will
help us to verify our understanding based on the predictions of
the standard model (SM), which has been very successful so far.
Predictions based on the SM remain almost unchallenged even today,
except possibly neutrino having a mass. On the other hand, the SM
contains many parameters which are unknown to a satisfactory level
of accuracy. Also the SM does not provide any explanation so as to
why there are only three generation of fermions, the mass
hierarchy among the fermions etc. There are many variants of
possible extensions to the SM exist in the literature and it is
widely believed that physics beyond the SM might be discovered
soon.

Unfortunately, we have not been able to see any indication of
physics beyond the SM in the currently running B-factories (SLAC
and KEK). Nevertheless, there appears to be some kind of deviation
in some $b\to s$ penguin induced transitions (like the deviation
in the measurement of sin2$\beta$ in $B_d \to \phi K_S $ and also in
some related processes \cite{zl1}, polarization anomaly in $B\to \phi K^*$
\cite{yg04} and deviation of branching ratios from the SM expectation 
in some rare B decays, etc.).
But, it is too early to claim or rule out the existence of new
physics in the b-sector. On the other hand, B-factories are expected
to continue accumulating data for some more years and thereafter
the task will be taken over by the new second generation
B-experiments such as LHC-b and BTeV. It is therefore perceived
very strongly that, in future, we will be able to identify the
existence of new physics (NP), if really there is any. It should
be mentioned here that one of the main objectives behind the
pursuance of B-factory experiments is to look for physics beyond
the SM.

One of the important ways to look for new physics in the b-sector is
the analysis of rare $B$ decay modes, which are induced by the
flavor changing neutral current (FCNC) transitions. The 
FCNC transitions generally arise at the loop level in the SM, 
thus provide an excellent
testing ground for new physics. Therefore, it is very important to
study the FCNC processes, both theoretically and experimentally,
as these decays can provide a sensitive test for the investigation
of the gauge structure of the SM at the loop level. Concerning the
semileptonic B decays, $B\to X_s~ l^+l^-$ ($X_s= K, K^*,~ l=e, \mu,
\tau$) are a class of decays having both theoretical and
experimental importance. At the quark level, these decays proceed
through the FCNC transition $ b\to s$, which occur only through
loops in the SM. For the very same reason, the study of the FCNC
decays can provide a sensitive test for the investigation of the
gauge structure of the SM at the loop level. At the same time,
these decays constitute a quite suitable tool of looking for new
physics. New physics effects manifest themselves in these rare $B$
decays in two different ways, either through new contribution to
the Wilson coefficients or through the new structure in the
effective Hamiltonian, which are absent in the SM.

It is well known that the theoretical analysis of the inclusive
decay is easy but their experimental detection is difficult. For
exclusive decays the situation is opposite, i.e., these decays can
be easily studied  in the experiments but theoretically they have
drawbacks and predictions are model dependent. This is due to the
fact that in calculating the branching ratios and other
observables for exclusive decays we face the problem of computing
the hadronic form factors.

Therefore, the exclusive processes induced by the quark level
transition $b\to s l^+l^-$ have received a considerable attention
in the literature because of its richness to study the FCNC.
Moreover, the dileptons present in these processes allow us to
formulate many observables which can serve as a testing ground to
decipher the presence of new physics. With the accumulation of
data, day by day in the b-sector, we are in an increasingly better
position to experimentally study the semileptonic decay induced by
$b\to s $ transition. In this context a rich and extensive study
exists \cite{sm} as far as the rare decay process $B\to K l^+l^-$ and its
vector counterpart $B\to K^* l^+l^-$ are concerned, in the
framework of the SM and in many extensions of it. However, the
study of baryonic rare semileptonic decay modes, also induced by
the same quark level transition, i.e.,  $b\to s l^+l^-$, are also
as important as its mesonic counterparts and deserve serious
attention, both theoretically and experimentally. Since at the
quark level they are induced by the same mechanism, so we can
independently test our understanding of the quark-hadron dynamics
and also study some CP violation parameters with the help of
baryonic rare decays, apart from corroborating the findings of the
mesonic sector.

In this work, we would like to analyze the rare baryonic decay
mode $\Lambda_b \to \Lambda~ l^+l^-$. We consider the effect of the
non-universal $Z$ boson which induces FCNC interaction at the tree
level. It is well known that FCNC coupling of the $Z$ boson can be
generated at the tree level in various exotic scenarios. Two
popular examples discussed in the literature are the models with
an extra $U(1)$ symmetry \cite{ref2} and those with the addition of
non-sequential generation of quarks \cite{ref3}. In the case of extra
$U(1)$ symmetry the FCNC couplings of the $Z$ boson are induced by
$Z-Z^\prime$ mixing, provided the SM quarks have family
non-universal charges under the new $U(1)$ group. In the second
case, adding a different number of up- and down-type quarks, the
pseudo CKM matrix needed to diagonalize the charged currents is no
longer unitary and this leads to tree level FCNC couplings. Here
we will follow the second approach to analyze the semileptonic
rare $\Lambda_b$  decays. These decays are studied in the SM
\cite{huang}, in the supersymmetric model with and without R-parity
\cite{chen} and in a model independent way by Aliev et al \cite
{aliev}. To have a complete understanding of the  nature of the
new physics, if it indeed exists, it would be worthwhile to analyze 
these rare decays in as many new physics models as possible. In this paper,
we would like to see the effect of the nonuniversal $Z$-boson in the decay
width and forward backward asymmetry of the lepton pairs, when the final
$\ll$ baryon is unpolarized. We also study its effect on the polarization
of $\ll$ baryon.

The paper is organized as follows. In section II we 
briefly describe the decay parameters of the semileptonic rare
decays in the Standard Model.
In section III the effect of the  FCNC mediated $Z$ boson has
been considered. The numerical results are
presented in section IV and section V contains the conclusion.

\section{Standard Model contribution}

The decay process $\lb \to \ll~ l^+ l^-$ is described by the 
quark level transition $ b \to s l^+ l^-$. Thus,  
the effective Hamiltonian describing this process can be given as
\cite{buras} 
\bea
{\cal H}_{eff} &= &\frac{ G_F~ \alpha}{\sqrt 2 \pi}~ V_{tb} V_{ts}^*~\Big[ 
C_9^{eff}(\bar s \gamma_\mu L b)(\bar l \gamma^\mu l) \nn\\
&+& C_{10}(\bar s \gamma_\mu L b)(\bar l \gamma^\mu \gamma_5 l) -2 C_7^{eff}
 m_b(\bar s i \sigma_{\mu \nu} \frac{q^\mu}{q^2} R b)
(\bar l \gamma^\mu l) \Big]\;,\label{ham}
\eea
where $q$ is the momentum transferred to the lepton pair, given as
$q=p_-+p_+$, with $p_-$ and $p_+$ are the momenta of the leptons $l^-$ 
and $l^+$ respectively. $L,R=(1 \pm \gamma_5)/2$ and $C_i$'s are the Wilson
coefficients evaluated at the $b$ quark mass scale. The values of these
coefficients in NLL order are \cite{beneke}
\be
C_7^{eff}=-0.308\;,~~C_9=4.154\;,~~C_{10}=-4.261\;.\label{wil}
\ee
The coefficient $C_9^{eff}$ has a perturbative part and a 
resonance part which comes
from the long distance effects due to the conversion of the real
 $c \bar c$ into the lepton pair $l^+ l^-$. Therefore, one can write it as 
\be
C_9^{eff}=C_9+Y(s)+C_9^{res}\;,
\ee
where $s=q^2$ and the function $Y(s)$ denotes the perturbative part coming 
from one loop matrix elements  of the four quark operators and
is given by \cite{buras1}
\bea
Y(s)&=& g(m_c,s)(3 C_1+C_2+3C_3+C_4+3C_5+C_6) -\frac{1}{2} g(0,s)
(C_3+3C_4)\nn\\
&-&\frac{1}{2} g(m_b,s)(4 C_3+4 C_4+3 C_5 +C_6)
+ \frac{2}{9}(3 C_3+C_4+3C_5+C_6)\;,
\eea
where
\bea
g(m_i,s) &=& -\frac{8}{9} \ln(m_i/m_b^{pole}) + \frac{8}{27}+\frac{
4}{9}y_i -\frac{2}{9}(2+y_i)\sqrt{|1-y_i|}\nn\\
&\times & \biggr\{\Theta(1-y_i)\biggr[\ln\left (
\frac{1+\sqrt{1-y_i}}{1-\sqrt{1-y_i}}\right )-i \pi \biggr]
+\Theta(y_i-1)2 \arctan \frac{1}{\sqrt{y_i-1}} \biggr\}\;,
\eea
with $y_i=4 m_i^2/s$. The values of the
coefficients $C_i$'s in NLL order are taken from \cite{beneke} as
$C_1=-0.151\;,~C_2=1.059\;,~C_3=0.012\;,~C_4=-0.034\;,~
C_5=0.010$ and $C_6=-0.040$.

The long distance resonance effect is given as \cite{res}
\bea
C_9^{res}= \frac{3 \pi}{\alpha^2}(3 C_1+C_2+3C_3+C_4+3C_5+C_6)\sum_{
V_i=\psi(1S), \cdots, \psi(6S) } \kappa_{V_i}\frac{m_{V_i} \Gamma(V_i \to l^+ l^-)}{m_{V_i}^2 -s
-i m_{V_i}\Gamma_{V_i}}\;.
\eea
The phenomenological parameter $\kappa$ is taken to be 2.3 so as to
reproduce the correct branching ratio of $ {\cal B}(B \to J/\psi K^*
\to K^* l^+ l^-)={\cal B}(B \to J/\psi K^*){\cal B}(J/\psi \to l^+ l^-)$.

After having an idea of the effective Hamiltonian and the relevant
Wilson coefficients, we now proceed to evaluate the transition matrix 
elements for the process $\lb (p_{\lb}) \to \ll (p_\ll)~
l^+ (p_+)l^-(p_-)$. For this purpose, we need to know the marix elements 
of the various hadronic 
currents between initial $\lb$ and the final $\ll$ baryon, which are 
parametrized in terms of various form factors as
\bea
\langle \ll |\bar s \gamma_\mu b | \lb \rangle & =& \bar u_\ll \Big[
f_1 \gamma_\mu + i f_2 \sigma_{\mu \nu} q^\nu +f_3 q_\mu \Big]
u_{\lb}\;,\nn\\
\langle \ll |\bar s \gamma_\mu \gamma_5 b | \lb \rangle & =& 
\bar u_\ll \Big[
g_1 \gamma_\mu \gamma_5 + i g_2 \sigma_{\mu \nu}\gamma_5 q^\nu +
g_3 \gamma_5 q_\mu \Big]
u_{\lb}\;,\nn\\
\langle \ll |\bar s i \sigma_{\mu \nu} q^\nu b | \lb \rangle & =&
\bar u_\ll \Big[
f_1^T \gamma_\mu + i f_2^T \sigma_{\mu \nu} q^\nu +f_3^T q_\mu \Big]
u_{\lb}\;,\nn\\
\langle \ll |\bar s i \sigma_{\mu \nu}\gamma_5 q^\nu b | \lb \rangle & =& 
\bar u_\ll \Big[
g_1^T \gamma_\mu \gamma_5 + i g_2^T \sigma_{\mu \nu}\gamma_5 q^\nu +
g_3^T \gamma_5 q_\mu \Big]
u_{\lb}\;,
\eea 
where $q=p_{\lb}-p_{\ll}=p_++p_-$ is the momentum transfer, $f_i$ and 
$g_i$ are the various form factors which are functions of $q^2$. 
The number of independent form factors are greatly reduced in the heavy
quark symmetry limit. In this  limit, the matrix elements of all
the hadronic currents, irrespective of their Dirac structure, 
can be given in terms of only two independent
form factors \cite{mannel} as
\be
\langle \ll(p_\ll) | \bar s \Gamma b | \lb (p_{\lb}) \rangle
= \bar u_\ll [F_1(q^2) +\not\!{v} F_2 (q^2) ] \Gamma u_{\lb}\;,
\ee
where $\Gamma$ is the product of Dirac matrices, $v^\mu=p_{\lb}^\mu/
m_{\lb}$ is the four velocity of $\Lambda_b$.
These two sets of form facors are relalated to each other as
\bea
&& g_1=f_1=f_2^T=g_2^T=F_1 +\sqrt r F_2\;,\nn\\
&& g_2=f_2=g_3=f_3=\frac{F_2}{m_{\lb}}\;,\nn\\
&&g_3^T=\frac{F_2}{m_{\lb}}(m_{\lb}+m_\ll)\;,~~~~~~
f_3^T=-\frac{F_2}{m_{\lb}}(m_{\lb}-m_\ll)\nn\\
&&f_1^T=g_1^T=\frac{F_2}{m_{\lb}}q^2\;,
\eea 
where $r=m_\ll^2/m_{\lb}^2$.
Thus, using these form factors,
the transition amplitude can be written as 
\bea
{\cal M}(\lb \to \ll l^+ l^-) &=& \frac{G_F~ \alpha}{\sqrt 2 \pi}
V_{tb} V_{ts}^* \Biggr[ \bar l \gamma_\mu l \Big\{
\bar u_\ll \Big(\gamma^\mu (A_1 P_R+B_1 P_L)+ i \sigma^{\mu \nu} q_\nu
(A_2 P_R +B_2 P_L) \Big)u_{\lb} \Big\}\nn\\
&+&\bar l \gamma_\mu \gamma_5 l \Big\{
\bar u_\ll \Big(\gamma^\mu  (D_1 P_R+E_1 P_L)+ i \sigma^{\mu \nu} q_\nu
(D_2 P_R +E_2 P_L)\nn\\
&+& q^\mu(D_3 P_R +E_3 P_L) \Big)u_{\lb} \Big\}\Biggr]\;,\label{e1}
\eea
where the various parameters $A_i,~B_i$ and $D_j,~E_j$ 
($i=1,2$ and $j=1,2,3$) are defined as
\bea 
A_i &=& \frac{1}{2}C_9^{eff}(f_i-g_i)-\frac{C_7 m_b}{q^2}
(f_i^T+g_i^T)\;,\nn\\
B_i &=& \frac{1}{2}C_9^{eff}(f_i+g_i)-\frac{C_7 m_b}{q^2}
(f_i^T-g_i^T)\;,\nn\\
D_j &=& \frac{1}{2}C_{10} (f_j-g_j)\;,~~~~
E_j = \frac{1}{2}C_{10} (f_j+g_j)\;.
\eea
Let us first consider the case when the final $\ll$ baryon is unpolarized.
The physical observables in this case are the differential decay rate and 
the forward backward rate asymmetries. From the transition 
amplitude (\ref{e1}), one can obtain
double differential decay rate as 
\bea
\frac{d^2 \Gamma}{d\hat s~ dz}=\frac{G_F^2~ \alpha^2}{2^{12} \pi^5}~
|V_{tb} V_{ts}^*|^2~m_{\lb}~ v_l~ \lambda^{1/2}(1, r, \hat s)~
{\cal K}(s ,z)\;, \label{e2}
\eea
where $\hat s=s/m_{\lb}^2$, $z=\cos \theta $, the angle
between $p_{\lb}$ and  $p_{+}$  in the center of mass
frame of $l^+ l^-$ pair, $v_l=\sqrt{1-4 m_l^2/ s}$
and $\lambda(a,b,c)=\sqrt{a^2+b^2+c^2-2(ab+bc+ca)}$ is the usual
triangle function.
The function ${\cal K}(s, z)$ is given as 
\be
{\cal K}(s,z)={\cal K}_0(s)+z~ {\cal K}_1(s)+z^2~{\cal K}_2(s)\;,
\ee
with
\bea
{\cal K}_0(s) &=& 32 m_l^2 m_{\lb}^2\s1(1+r -\s1)(|D_3|^2+|E_3|^2)\nn\\
&+&
64 m_l^2 m_{\lb}^3(1-r -\s1)Re(D_1^*E_3+D_3 E_1^*)
+64  m_{\lb}^2 \sqrt{r} (6 m_l^2-\s1 m_{\lb}^2)Re(D_1^*E_1)
\nn\\
&+&64 m_l^2 m_{\lb}^3 \sqrt{r} \Big(
2 m_{\lb} \s1 Re(D_3^*E_3)+(1-r+\s1)Re(D_1^* D_3+ E_1^* E_3)\Big)\nn\\
&+&32 m_{\lb}^2 (2 m_l^2+m_{\lb}^2 \s1)\Big((1-r +\s1)m_{\lb}
\sqrt{r}Re(A_1^*A_2+B_1^* B_2)\nn\\
&-& m_{\lb}(1-r-\s1) Re(A_1^* B_2+A_2^* B_1)-2 \sqrt{r}\Big[Re(A_1^* B_1)
+m_{\lb}^2 \s1 Re(A_2^* B_2)\Big]\Big)\nn\\
&+& 8 m_{\lb}^2\Big(4 m_l^2(1+r-\s1)+m_{\lb}^2[(1-r)^2-\s1^2]\Big)
\Big(|A_1|^2+|B_1|^2\Big)\nn\\
&+& 8 m_{\lb}^4\Big(4 m_l^2[\lambda+(1+r-\s1)\s1]+m_{\lb}^2
\s1[(1-r)^2-\s1^2]\Big)
\Big(|A_2|^2+|B_2|^2\Big)\nn\\
&-& 8 m_{\lb}^2\Big(4 m_l^2(1+r-\s1)-m_{\lb}^2[(1-r)^2-\s1^2]\Big)
\Big(|D_1|^2+|E_1|^2\Big)\nn\\
&+& 8 m_{\lb}^5 \s1 v_l^2 \Big(-8 m_{\lb} \s1 \sqrt{r}
Re(D_2^* E_2)+4 (1-r+\s1)\sqrt{r}Re(D_1^* D_2+E_1^* E_2)\nn\\
&-&4(1-r -\s1) Re(D_1^* E_2+D_2^* E_1)
+m_{\lb}[(1-r)^2-\s1^2]
\Big[|D_2|^2+|E_2|^2\Big]\Big)\;,
\eea
\bea
{\cal K}_1(s) &=& -16  m_{\lb}^4\s1 v_l \sqrt{\lambda}
\Big\{ 2 Re(A_1^* D_1)-2Re(B_1^* E_1)\nn\\
&+& 2m_{\lb}
Re(B_1^* D_2-B_2^* D_1+A_2^* E_1-A_1^*E_2)\Big\}\nn\\
&+&32 m_{\lb}^5 \s1~ v_l \sqrt{\lambda} \Big\{
m_{\lb} (1-r)Re(A_2^* D_2 -B_2^* E_2)\nn\\
&+&
\sqrt{r} Re(A_2^* D_1+A_1^* D_2-B_2^*E_1-B_1^* E_2)\Big\}\;,
\eea
and
\bea
{\cal K}_2(s)&= & 8m_{\lb}^6 v_l^2~ \lambda \s1~ \Big ( 
(|A_2|^2+|B_2|^2+|D_2|^2+|E_2|^2\Big)\nn\\
&-&8m_{\lb}^4 v_l^2 ~\lambda~\Big(|A_1|^2+|B_1|^2+|D_1|^2+|E_1|^2\Big)\;.
\eea
The dilepton mass spectrum can be obtained from (\ref{e2}) by integrating
out the angular dependent parameter $z$ which yields
\be
\left (\frac{d \Gamma}{d s}\right )_0= \frac{G_F^2~ \alpha^2}
{2^{11} \pi^5 m_{\lb}}~
|V_{tb}V_{ts}^*|^2 v_l~ \sqrt{\lambda}~\Big[{\cal K}_0(s)+ \frac{1}{3}
{\cal K}_2(s)\Big]\;,\label{dl}
\ee
where $\lambda$ is the short hand notation for
$\lambda(1, r, \hat s)$. The limits for $s$ is
\be
4 m_l^2 \leq s\leq (m_{\lb}-m_\ll)^2\;.
\ee

Another observable is the lepton forward backward asymmetry ($A_{FB}$),
which is also a very powerful tool for looking  new physics. The
position of the zero value of $A_{FB}$ is very sensitive to
the presence of new physics.

The normalized forward-backward asymmetry
is defined as
\be
A_{FB}(s)=\frac{\ds{\int_0^1 \frac{d \Gamma}{d \s1 dz}dz-\int_{-1}^0 
\frac{d \Gamma}{d \s1 dz}dz}}
{\ds{\int_0^1 \frac{d \Gamma}{d \s1 dz}dz+\int_{-1}^0 
\frac{d \Gamma}{d \s1 dz}dz}}\;.
\ee
Thus one obtains from (\ref{e2}) 
\be
A_{FB}(s)=\frac{{\cal K}_1(s)}{{\cal K}_0(s)+{\cal K}_2(s)/3}\;.\label{fb}
\ee
Now let us consider the case when the final $\ll$ baryon is
polarized. To study its spin polarization, we
write the spin  vector of $\ll$  in terms of a unit vector 
$\hat \eta$ along 
the direction of $\ll$ spin, in its rest frame as
\be
s_\mu= \Big(\frac{\vec p_\ll \cdot \hat \eta}{m_\ll}\;,~~ \hat \eta+   
\frac{\vec p_\ll \cdot \hat \eta}{m_\ll(E_\ll+m_\ll)}\vec p_\ll \Big)\;.
\ee
We also consider  three orthogonal  unit vectors along the 
longitudinal, transverse and normal
components of $\ll$ polarization in the $\lb$ rest frame as
\be
\hat e_L= \frac{\vec p_\ll}{|\vec p_\ll|}\;,~~~
\hat e_T= \frac{\vec p_+ \times \vec p_\ll }{|\vec p_+ \times
\vec p_\ll|}\;,~~~~\hat e_N=\hat e_T \times \hat e_L\;,
\ee
where $\vec p_\ll$ and $\vec p_+$ are three momenta of the $\ll$ and $l^+$
in the c.m. frame of the $l^+ l^-$ system.
Thus, using these spin vectors one can 
obtain the differential decay rate for any spin direction $\hat \eta $
along the $\ll$ baryon  as
\be
\frac{d \Gamma(\hat \eta)}{d s}=
\frac{1}{2}\left (\frac{d \Gamma}{d s} \right )_0\Big[
1+\left ( P_L ~\hat e_L + P_N ~\hat e_N +P_T~ \hat e_T \right )\cdot \hat \eta
\Big]\;,\label{dl1}
\ee
where $P_L$, $P_N$ and $P_T$ are functions of $s$, which give the 
longitudinal, normal and transverse polarization and $\left (d \Gamma/
d s \right )_0$ is the unpolarized decay width.
The polarization components $P_i$ ($i=L, N, T$) can be obtained from
\be
P_i(s)= \frac{\ds{\frac{d \Gamma}{ds}(\hat \eta= \hat e_i)-
\frac{d \Gamma}{ds}(\hat \eta= -\hat e_i)}}
{\ds{\frac{d \Gamma}{ds}(\hat \eta= \hat e_i)+
\frac{d \Gamma}{ds}(\hat \eta= -\hat e_i)}}\;.
\ee
Thus, one can obtain the polarization components as
\bea
P_L(s) &=& \frac{16 m_{\lb}^2 \sqrt{\lambda}}{
{\cal K}_0(s)+{\cal K}_2(s)/3}\biggr[ 8 m_l^2 m_{\lb}\Big
( Re(D_1^* E_3 -D_3^* E_1)+
\sqrt{r} Re(D_1^* D_3 -E_1^* E_3) \Big) \nn\\
&-& 4 m_l^2 m_{\lb}^2 \s1 \Big(|D_3|^2-|E_3|^2
\Big) -4 m_{\lb}(2 m_l^2+m_{\lb}^2
\s1)Re(A_1^* B_2-A_2^* B_1)\nn\\
&-& \frac{4}{3} m_{\lb}^3 \s1~ v_l^2\Big(3 Re(D_1^* E_2-D_2^* E_1)
+\sqrt{r} Re(D_1^* D_2-E_1^* E_2)\Big)\nn\\
&-& \frac{4}{3}m_{\lb} \sqrt{r} (6 m_l^2 +m_{\lb}^2 \s1~ v_l^2)
Re(A_1^* A_2-B_1^* B_2)-\frac{2}{3} m_{\lb}^4 \s1 (2-2r+\s1) 
v_l^2 (|D_2|^2-|E_2|^2)\nn\\
&+& (4 m_l^2+m_{\lb}^2(1-r+\s1))(|A_1|^2-|B_1|^2)
-(4m_l^2-m_{\lb}^2(1-r+\s1))(|D_1|^2-|E_1|^2)\nn\\
&-&\frac{1}{3} m_{\lb}^2(1-r-\s1)~ v_l^2~ (|A_1|^2-|B_1|^2+|D_1|^2-
|E_1|^2)\nn\\
&-& \frac{1}{3} m_{\lb}^2\Big[12 m_l^2(1-r)+m_{\lb}^2 \s1(3(1-r+\s1)
+v_l^2 (1-r-\s1))\Big]\Big(|A_2|^2-|B_2|^2)\biggr]
\eea
\bea
P_N(s) &=& \frac{8 \pi m_{\lb}^3 v_l \sqrt{\s1}}{
{\cal K}_0(s)+{\cal K}_2(s)/3}\biggr[-2 m_{\lb}(1-r+\s1)\sqrt{r}
Re(A_1^* D_1+B_1^* E_1)\nn\\
&+& 4 m_{\lb}^2 \s1 \sqrt{r} Re(A_1^* E_2+A_2^* E_1+B_1^* D_2
+B_2^* D_1)\nn\\
&-& 2 m_{\lb}^3 \s1 \sqrt{r} (1-r+\s1)Re(A_2^* D_2+B_2^* E_2)\nn\\
&+& 2 m_{\lb} (1-r-\s1)\Big( Re(A_1^* E_1+B_1^* D_1)+m_{\lb}^2 
\s1 Re(A_2^* E_2+ B_2^* D_2)\Big)\nn\\
&-& m_{\lb}^2\Big((1-r)^2-\s1^2\Big)Re(A_1^* D_2 +A_2^* D_1+B_1^* 
E_2 +B_2^* E_1)
\biggr]\;,\label{pn}
\eea
\bea
P_T(s) &=& - \frac{8 \pi m_{\lb}^3 v_l \sqrt{\s1 \lambda}}{{\cal K}_0(s)+ 
{\cal K}_2 (s)/3}\biggr[ m_{\lb}^2(1-r+\s1)\Big(
Im(A_2^* D_1-A_1^* D_2)\nn\\
&-&Im(B_2^* E_1-B_1^* E_2)\Big)
+ 2 m_{\lb}\Big(Im(A_1^*E_1-B_1^* D_1)\nn\\
&-&m_{\lb}^2 \s1 
Im(A_2^* E_2-B_2^* D_2)\Big)
\biggr]\;.
\eea
It should be noted that the longitudinal ($P_L$) and normal ($P_N$) 
polarizations are P-odd and T-even whereas the transverse
polarization ($P_T$) is P-even and T-odd.
 
\section{Contribution from FCNC mediated $Z$ boson}

We now consider the effect of FCNC mediated $Z$ boson on the branching
ratios and forward backward asymmetries of the rare semileptonic
$\lb$ decays. It is a simple model beyond
the standard model with an enlarged matter sector due to an additional
vector like down quark $D_4$. The presence of an additional
down quark implies a $4 \times 4$ matrix $V_{i \alpha}$ $(i=u,c,t,4,
 ~ \alpha=d,s,b, b^\prime)$, diagonalizing the down quark
mass matrix. For our purpose the relevant information for
the low energy physics is encoded in the extended mixing matrix.
The charged currents are unchanged except that the $V_{CKM}$ is now
the $3 \times 4$ upper sub-matrix of $V$. However, the distinctive
feature of this model is that the FCNC interaction enters neutral current
Lagrangian of the left handed down quarks as
\be
{\cal L}_Z= \frac{g}{2 \cos \theta_W}\Big[\bar u_{Li}
\gamma^\mu u_{Li}-\bar d_{L \alpha}U_{\alpha \beta} \gamma^\mu
d_{L \beta}-2 \sin^2 \theta_W J^\mu_{em}\Big] Z_\mu\;,
\ee
with
\be
U_{\alpha \beta}= \sum_{i=u,c,t} V_{\alpha i}^\dagger V_{i \beta}=
\delta_{\alpha \beta}-V_{4 \alpha}^* V_{4 \beta}\;,
\ee
where $U$ is the neutral current mixing matrix for the
down sector, which is given above. As $V$ is not unitary,
$U \neq {\bf 1}$. In particular the non-diagonal
elements do not vanish.
\be
U_{\alpha \beta}=-V_{4 \alpha}^* V_{4 \beta} \neq 0 ~~~~
{\rm for}~~~\alpha \neq \beta\;.
\ee
Since the various $U_{\alpha \beta}$ are non vanishing, they would
signal new physics and the presence of FCNC at the tree level and this can
substantially modify the predictions of SM for the FCNC processes.

Thus, in this model the effective Hamiltonian for 
$b \to s l^+ l^-$ is given as
\be
{\cal H}_{eff}= \frac{G_F}{\sqrt 2}~ U_{sb}~ [\bar s \gamma^\mu
(1-\gamma_5)b]
\left [\bar l( C_V^l \gamma_\mu  -C_A^l \gamma_\mu \gamma_5 ) l
\right ]\;,\label{ham1}
\ee
where $C_V^l$ and $C_A^l$  are the vector and axial vector $Z l^+ l^-$
couplings, which are given as
\be
C_V^l= -\frac{1}{2}+2\sin^2 \theta_W\;,~~~~~~~
C_A^l= -\frac{1}{2}\;. \label{ca}
\ee
Since, the structure of the effective Hamiltonian (\ref{ham1}) in
this model is same as that of the SM, i.e., $\sim(V-A)(V-A)$ form, its 
 effect on the various decay parameters can be encoded by replacing the 
SM Wilson coeiffients $(C_9^{eff})^{SM}$ and $(C_{10})^{SM}$ by 
\bea
&&C_9^{eff} =( C_9^{eff})^{SM}+\frac{2 \pi}{\alpha}\frac{ U_{sb}}{
V_{tb}V_{ts}^*}\nn\\
&&
C_{10}^{eff} = (C_{10})^{SM}-\frac{2 \pi}{\alpha}\frac{ U_{sb}}{
V_{tb}V_{ts}^*}\;.
\eea

It should be noted that $U_{sb}$ is in general complex and hence  it
induces the weak phase difference ($\theta$) between the SM and new 
physics contributions. Since the value of the Wilson coefficients $C_9$ 
and $C_{10}$ are opposite to each other as seen from Eq. (\ref{wil}),
and  the  new physics contributions to $C_9$ and $C_{10}$ are opposite 
to each other, one will get constructive or destructive interference 
of SM and NP amplitudes for $\theta=\pi$ or zero (where $\theta$ denotes the
relative weak phase between SM and NP contribution in the above equation). 
However, we consider the
weak phase difference to be $\pi$  in our numerical analysis, so as to 
get the constructive interference between the SM and NP amplitudes.
The value of $|U_{sb}|$ is found to be 
\be
|U_{sb}| \simeq 10^{-3}\;,
\ee
which has been
extracted from the recent data on ${\cal B}(B \to X_S l^+
l^-)$ \cite{ref6}. 

\section{Numerical Analysis}

For numerical evaluation we use the various particle masses and
lifetimes of $\lb$ baryon from \cite{pdg}. The quark masses
(in GeV) used are
$m_b$=4.6, $m_c$=1.5, the CKM matrix elements as
$|V_{tb} V_{ts}^*|=0.041$, $\alpha=1/128$ and the weak mixing angle 
$\sin^2 \theta_W=0.23$.
For the form factors we use the values calculated in the QCD sum rule 
\cite{huang,chen}
approach, where the $q^2$ dependence of various form factors are given as
\be
F(q^2)=\frac{F(0)}{1-a_F (q^2/m_{\lb}^2)+b_F \left ( q^2/m_{\lb}^2
\right )^2}\;.
\ee
The values of the parameters $F_i(0)$, $a$ and $b$  are summarized in 
Table-1. 
\begin{table}
\begin{center}
\caption{Values of the hadronic form factors in the QCD sum rule approach for
$\lb \to \ll$ transition.}
\vspace*{0.3 true cm}
\begin{tabular}{cccccccccc}
\hline
\hline
Form Factors &&& $F(0)$ &&& $a_F$ &&& $b_F$\\
\hline
\hline
$F_1$ &&& 0.462 &&& $-0.0182$ &&& $-1.76 \times 10^{-4}$\\
$F_2$ &&& $-0.077$  &&& $-0.0685$ &&& $1.46 \times 10^{-3}$\\
\hline
\hline
\end{tabular}
\end{center}
\end{table}

\begin{table}
\begin{center}
\caption{The branching ratios (in units of $10^{-6}$)
for various decay processes.}
\vspace*{0.3 true cm}
\begin{tabular}{ccccccccc}
\hline
\hline
Decay modes &&&&  $ {\cal B}^{SM} $ &&&& $ {\cal B}^{NP} 
$  \\
\hline
\hline
$\lb \to \ll \mu^+ \mu^- $ &&&& 4.55  &&&& 20.9 \\
$\lb \to \ll \tau^+ \tau^- $ &&&& 0.17 &&&& 0.92  \\
\hline
\hline
\end{tabular}
\end{center}
\end{table}

With these values we plot the differential decay rate (\ref{dl}) for
$\lb \to \ll~ l^+ l^- $ for $l=\mu,\tau$,  against $s$, which are
depicted in Figures-1 and 2. It can be noted from the figures that
there is considerable enhancement in the decay rates due to the
non-universal $Zbs$ coupling. The forward backward asymmetries
(\ref{fb}) are plotted  in Figures-3 and 4. It can be seen from
Figure-3 that the zero position of 
$A_{FB}$ shifts towards the left for $\lb \to \ll \mu^+ \mu^-$
process due to the NP effect, however there is no such deviation
in the $ \lb \to \ll \tau^+ \tau^-$ process. The differential 
decay rate for the longitudinal polarized $\ll$ are shown
in Figs.-5 and 6. These distiributions are very similar to the
differential decay rates (Figures-1 and 2) but with opposite sign.
We also find that there is no significant difference in the longitudinal 
polarization of $\ll$ due to the NP effect. The distribution of the
normal polarization components are shown in Figs.-7 and 8. From these
figures one can observe that the normal polarization is very small
in the region with low momentum transfer and can have significant
values in the large momentum transfer region. The transverse polarization
$P_T$ is found to be identically zero in this model as the
structure of the Hamiltonian is same as that of the SM, i.e., 
$(V-A)(V-A)$ form.
\begin{figure}[htb]
  \centerline{\epsfysize 2.25 truein \epsfbox{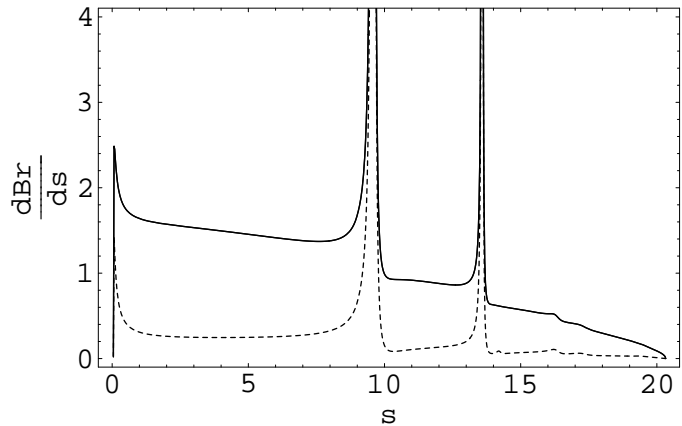}}
 \caption{
  The differential branching ratio $d {\rm Br}/d s$ (in units of $10^6~
{\rm GeV^{-2}}$)
versus $s~({\rm in~GeV^2})$
for the process  $\lb \to  \ll ~\mu^+ \mu^- $. The solid line denotes the 
branching ratio including the non-universal $Z$-boson effect 
whereas the dashed 
line represents the SM contribution.}
  \end{figure}

\begin{figure}[htb]
   \centerline{\epsfysize 2.25 truein \epsfbox{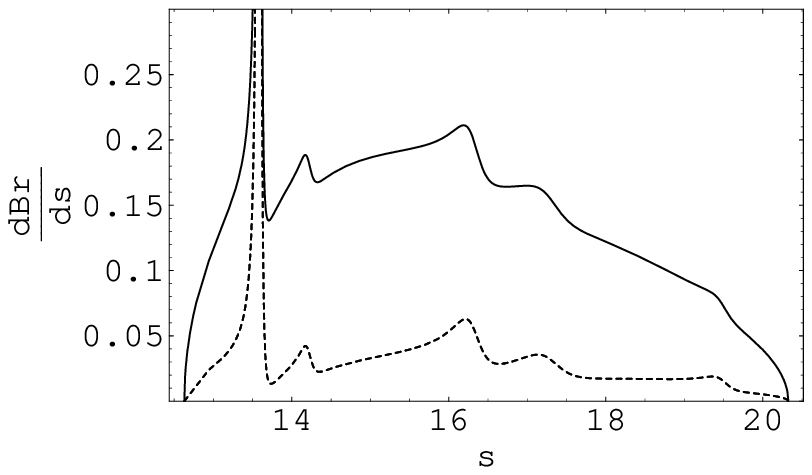}}
 \caption{Same as Figure-1 for the  $\lb \to  \ll~ \tau^+ \tau^- $
process. }
  \end{figure}

\begin{figure}[htb]
   \centerline{\epsfysize 2.25 truein \epsfbox{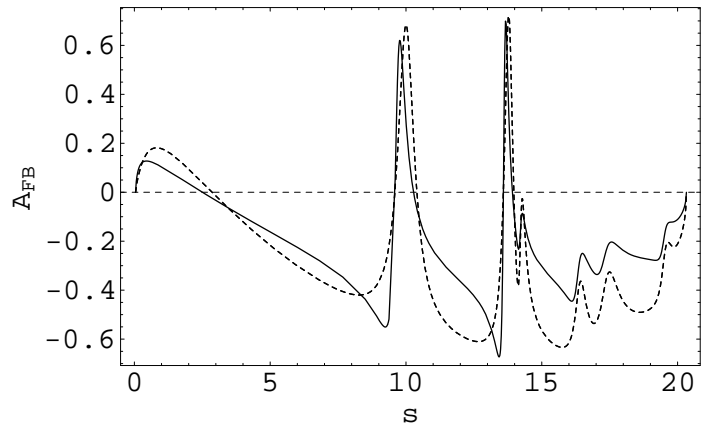}}
 \caption{The Forward-backward asymmetry vs. s (in ${\rm GeV^2}$ 
for the process 
$\lb \to \ll~ \mu^+ \mu^-$. The legends are same as Figure-1.
  }
\end{figure}

\begin{figure}[htb]
   \centerline{\epsfysize 2.25 truein \epsfbox{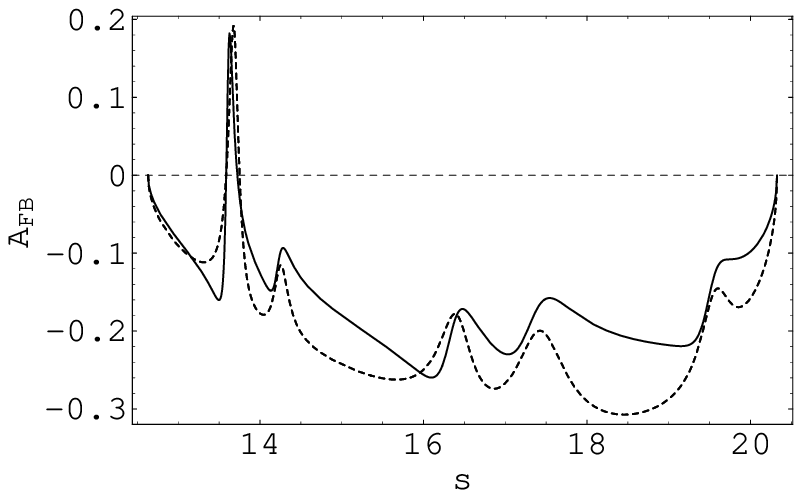}}
 \caption{Same as Figure-3 for the  $\lb \to  \ll~ \tau^+ \tau^- $
process.
  }
\end{figure}
\begin{figure}[htb]
  \centerline{\epsfysize 2.25 truein \epsfbox{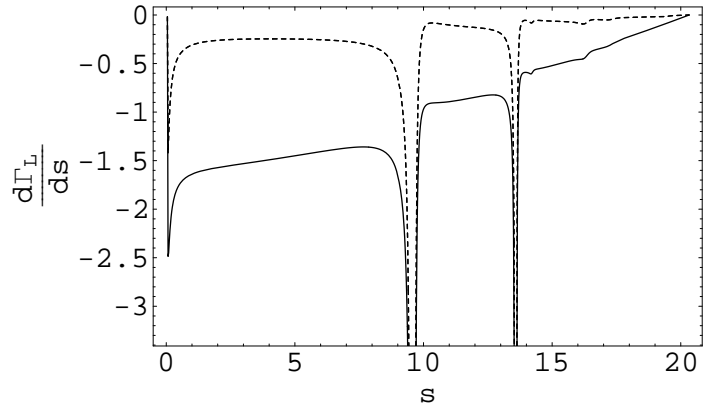}}
 \caption{
  The differential decay width distribution $d {\Gamma}_L/d s $
(in units of $\tau_{\lb}\times 10^6~{\rm GeV^{-2}}$)
versus $s~({\rm in~GeV^2})$
of $\lb \to \ll~ \mu^+ \mu^- $ for the 
longitudinal polarized $\ll$. The legends are same as Figure-1.}
  \end{figure}

\begin{figure}[htb]
   \centerline{\epsfysize 2.25 truein \epsfbox{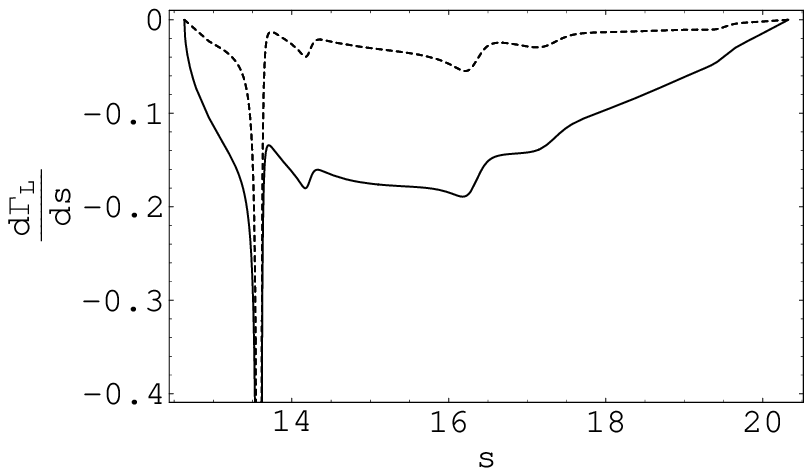}}
 \caption{Same as Figure-5 for the  $\lb \to  \ll~ \tau^+ \tau^- $
process. }
  \end{figure}

\begin{figure}[htb]
   \centerline{\epsfysize 2.25 truein \epsfbox{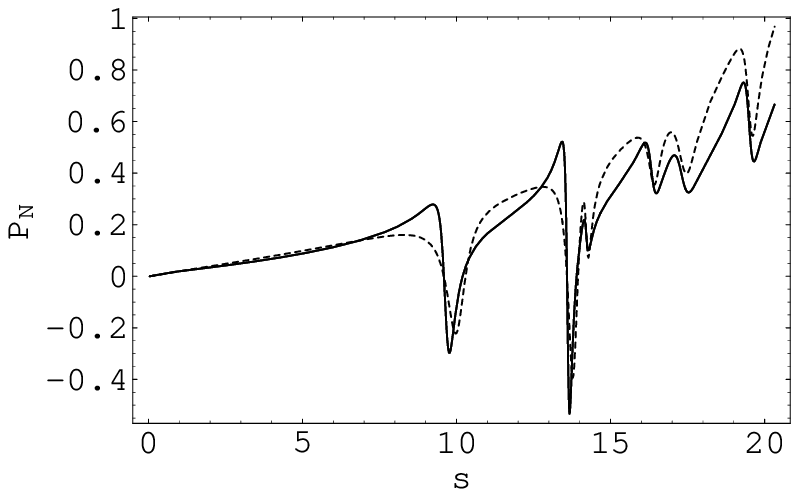}}
 \caption{Normal polarization $P_N$ vs. s (in GeV$^2$)
 for the  $\lb \to  \ll~ \mu^+ \mu^- $
process.
  }
\end{figure}

\begin{figure}[htb]
   \centerline{\epsfysize 2.25 truein \epsfbox{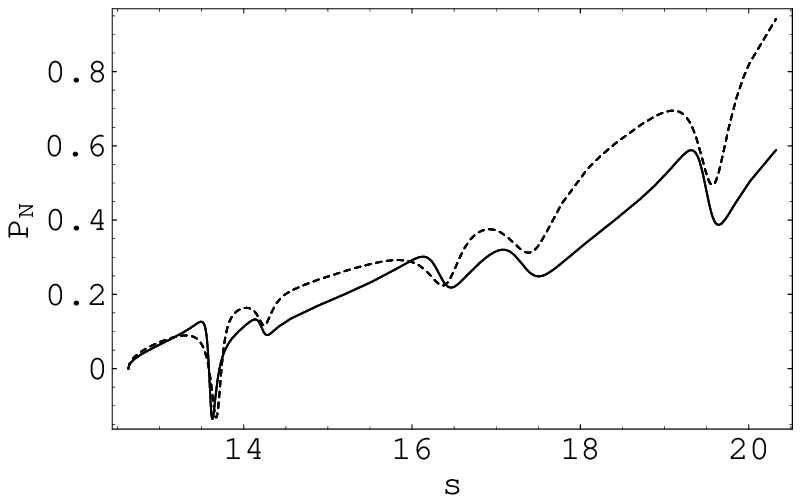}}
 \caption{Same as Figure-7 for the  $\lb \to  \ll~ \tau^+ \tau^- $
process.
  }
\end{figure}

We now proceed to calculate the  total decay rates for $\lb \to
\ll~ l^+ l^-$ for which it is necessary to eliminate the backgrounds
coming from the resonance regions. This can be done by by using the
follwing veto windows so that the backgrounds coming from the
dominant resonances $\lb \to \ll J/\psi (\psi^\prime)$ with
$J/\psi(\psi^\prime)\to l^+ l^-$ can be elimenated, 
\begin{eqnarray*}
\lb \to \ll~ \mu^+ \mu^-:&&m_{J/\psi}-0.02<
m_{\mu^+ \mu^-}<m_{J/\psi}+0.02;\nn\\
:&&
 m_{\psi^\prime}-0.02<m_{\mu^+ \mu^-}<m_{\psi^\prime}+0.02 \nn\\
\lb \to  \ll~ \tau^+ \tau^-:&&
m_{\psi^\prime}-0.02<m_{\tau^+ \tau^-}<m_{\psi^\prime}+0.02 \;.
\end{eqnarray*}
Using these veto windows we obtain the branching ratios for semileptonic
rare $\lb$ decays which are presented in Table-2. It is seen from the
table that the branching ratios obtained in the model with the
non-universal $Z$ bosons are reasonably enhanced from the 
corresponding SM values.

\section{Conclusion}

In this paper we have studied the rare baryonic semileptonic decay
$\Lambda_b \to \Lambda l^+l^-$ in the model in which the
fermion sector of the SM is extended by an extra vector like down quark.
The importance of this model is that it allows FCNC transitions
at the tree level. For the process under consideration,
it exhibits $b \to s$ FCNC transition at the
tree level by emitting one $Z$ boson. Here we have studied the effect
of this non universal
$Zbs$ couplings on the decay rates and  forward 
backward asymmetries of $\lb \to \ll~l^+ l^-$ process. Furthermore, we have 
also studied 
the diffrential decay rates when the final
$\ll$ baryon is longitudinally polarized and  the normal polarization
of the final $\ll$ baryon.
We found that in this model the branching ratios of $\lb \to \ll~ l^+ l^-$
can differ significantly from their corresponding SM values.
The forward backward asymmetries are also found to differ from that of the
SM expectation due to this non-universal $Z$ mediated FCNC. Moreover,
in this model the zero-point of the $F_{AB}$ is found to be shifted
towards left. For the polarized $\ll$, we found that the decay distribution
is similar to that of the unpolarized one but with opposite
orientation. Furthermore, no significant change in $P_N$ is observed,
$P_T$ is found to be identically zero.

To conclude, we note that the effect of $Z$ mediated FCNC in the vector
like down quark model can enhance the branching ratio reasonably
for the $\lb \to \ll l^+ l^-$ and also change the forward backward
asymmetries in these modes. The polarized variables can be studied
experimentally to distinguish various new physics models. The
transverse polarization $(P_T)$ in $\lb \to \ll~ l^+ l^-$, if found
zero (or nonzero) can confirm or (rule out) this new physics
scenario.

{\bf Acknowledgments}
AKG was supported by the National Science Council of the Republic
of China under Contract Number NSC-93-2811-M-007-060.
The work of RM was
partly supported by the Department of Science and Technology,
Government of India, through Grant No. SR/FTP/PS-50/2001.


\end{document}